\documentclass[aps,prd,floatfix,showpacs,nofootinbib,twoside,notitlepage,11pt]{revtex4-1}
\usepackage{amssymb}
\usepackage{amsmath}
\usepackage{amstext}
\usepackage{graphicx,epsfig}
\usepackage{verbatim} 
\usepackage{fancybox}
\usepackage{color}
\usepackage{ulem}
\usepackage{enumitem}
\usepackage{subfigure}
\usepackage{mwe}
\usepackage{bbm}
\usepackage{dsfont}
\usepackage{amsmath}		   
\usepackage{amssymb}		   
\usepackage{graphicx} 		   
\usepackage[sort&compress]{natbib} 
\usepackage{hyperref} 		   
\hypersetup{
   	colorlinks=true,	   
	linkcolor=blue,		   
 	citecolor=blue,		   
    	urlcolor=blue		   
	    }
\newcommand{\Mol}{\text{\rm M\o l}}
\newcommand{\Mpl}{M_{\rm pl}}
\newcommand{\MeV}{\ \mathrm{MeV}}
\newcommand{\cm}{\ \mathrm{cm}}
\newcommand{\s}{\ \mathrm{s}}

\newcommand{\eref}[1]{Eq.~(\ref{#1})}

\newcommand{\fref}[1]{Fig.~\ref{#1}}
\newcommand{\per}{\ . \ }
\newcommand{\com}{\ , \ }
\newcommand{\EQ}{\text{\sc EQ}}

\begin{document}

\setlength{\pdfpageheight}{\paperheight}
\setlength{\pdfpagewidth}{\paperwidth}


\title{Dibaryons cannot be the dark matter}

\author{Edward W. Kolb and Michael S. Turner}

\affiliation{Kavli Institute for Cosmological Physics and the Enrico Fermi Institute, The University of Chicago, 5640 S. Ellis Ave, Chicago, Il 60637}

\begin{abstract}
The  hypothetical $SU(3)$ flavor-singlet dibaryon state $S$ with strangeness $-2$ has been discussed as a dark-matter candidate capable of explaining the curious 5-to-1 ratio of the mass density of dark matter to that of baryons.  We study the early-universe production of dibaryons and find that irrespective of the hadron abundances produced by the QCD quark/hadron transition, rapid particle reactions thermalized the $S$ abundance, and it tracked equilibrium until it ``froze out'' at a tiny value.  For the plausible range of dibaryon masses ($1860 - 1890 \MeV$) and generous assumptions about its interaction cross sections, $S$'s account for at most $10^{-11}$ of the baryon number, and thus cannot be the dark matter.  Although it is not the dark matter, if the $S$ exists it might be an interesting relic.
\end{abstract}


\date{\today}

\maketitle

\section{Introduction}\label{sec:Introduction}
The most important thing we know about dark matter is that it is non-baryonic,\footnote{To be precise, by ``non-baryonic'' we mean that the dark matter was not in form of atoms, nuclei, or nucleons after the time of big-bang nucleosynthesis ($t \ge 1\,$sec).} and thus a new kind of matter.  This links the fields of cosmology and particle physics, and today, the nature of dark matter is one of the most pressing problems in each field.  The evidence for this comes from the primordial abundance of deuterium and the CMB angular power spectrum. In particular, there is a 50$\sigma$ discrepancy between the baryon density, $\Omega_Bh^2 = 0.0222 \pm 0.0002$, and the total matter density, $\Omega_Mh^2 = 0.142 \pm 0.0013$, inferred from BBN and the CMB; see e.g., Ref.\ \cite{Aghanim:2018eyx}.  

This fact also leads to a puzzle:  the unchanging ratio between the dark-matter density and the density of baryons is about 5-to-1, close to unity, rather than being very small or very large \cite{Turner:1987pp}.  The two leading candidates for dark matter, WIMP's and axions, may be compelling from the particle physics point-of-view, but do not address the 5-to-1 ratio in a compelling way.  One dark-matter candidate that did address this issue was quark nuggets \cite{Witten:1984rs}, where the dark matter was supposed to exist in the form of macroscopic, stable quark states with very large baryon number, which formed in a first-order QCD phase transition.  In this scenario, the order unity ratio of dark matter (quark nuggets) to baryons can arise naturally.

Our understanding of QCD and the quark/hadron transition now strongly disfavors this idea (but see Ref.\ \cite{Bai:2018vik}).  However, Farrar \cite{Farrar:2017eqq,Farrar:2018hac} has revived a variant of this idea with a stable (or very long-lived) dibaryon:  she argues that the $S$ dibaryon,\footnote{The $S$ dibaryon is sometimes referred to as the exaquark, or the sexaquark.} a six-quark configuration of 2 up quarks, 2 down quarks, and 2 strange quarks, an $SU(3)$ flavor-singlet with baryon number $2$, strangeness $-2$, and spin zero, could explain the ratio of dark matter to baryons.

To date there is no experimental evidence for the $S$ dibaryon, and its existence and properties are a topic of continuing debate in the QCD community.  Moreover, there are significant constraints from lattice QCD, hypernuclei, the ALICE experiment at the LHC, and unsuccessful searches \cite{Gongyo:2017fjb,Inoue:2010es,Acharya:2018gyz,Takahashi:2001nm}.  In this paper, we focus on the viability of the dibaryon dark-matter hypothesis, and not the existence of the dibaryon itself.  Fortunately, what we know with certainty about the putative $S$, summarized below, is sufficient to determine its cosmological production:
\begin{enumerate}
\item Its mass must be greater than $m_p + m_n - m_e -2BE \simeq 1860 \MeV$ ($BE \sim 8\MeV$ is the binding energy of a nucleon in a nucleus) to guarantee nuclear stability \cite{Farrar:2018hac}.\footnote{This constraint ensures that it is energetically impossible for a nucleus to decay into a nucleus with one less proton and one less neutron and an $S$.  We believe there is a much stronger limit based upon the stability of the deuteron.  If $m_S < m_p + m_n - m_e - 2.22 \MeV \simeq 1875\MeV$, the deuteron can decay via a doubly weak process into an $S$, a positron and a neutrino, with decay width $\Gamma > G_F^4 \delta m^9$.  The deuterons in the universe today were produced in the big bang 14\,Gyr ago; stability of the deuteron for the age of the universe requires $\delta m < 3\MeV$ or $m_S > 1872\MeV$.  Stronger constraints likely follow from the formation of neutron stars \cite{McDermott:inprep} and observed neutron-star masses \cite{BaymTurner:inprep}.}
\item  If its mass is greater than $2m_p + 2m_e \simeq 1878 \MeV$ it can decay to two nucleons; for $m_S < 2055\MeV = m_p + m_e + m_\Lambda$, this is a doubly-weak process with a decay width $\Gamma > G_F^{4} \delta m^9$ ($\delta m = m_S - 1878\MeV$);  for $m_S < 1890\MeV$ the lifetime of the $S$ is greater than the age of the universe.
\item  For an $S$ mass between $2055\MeV$ and $2232\MeV$ ($2232\MeV=2 m_\Lambda$) the dibaryon can decay via a singly weak process into a nucleon and a $\Lambda$ with a lifetime short compared to the age of the universe ($\Gamma > G_F^2 \delta m^5$), and for $m_S > 2232\MeV$ the dibaryon can decay into two $\Lambda$'s via the strong interaction with a very short lifetime.
\item In general, the non-strangeness changing interactions of the $S$, e.g., $\Lambda + \Lambda \longleftrightarrow \pi + \pi +S$, should be strong, although they might be suppressed by factors arising from the wavefunctions of the states involved \cite{Farrar:2003qy,Farrar:2003is}.
\end{enumerate}
These facts point to the  mass range 1860 MeV to 1890 MeV, where the dibaryon is stable or long-lived, and indicate that the reactions that control its abundance are strong and involve lambdas and possibly other strange baryons (e.g., strangeness $-1$ $\Sigma$'s and strangeness $-2$ $\Xi$'s).

The starting point for our analysis is the early-universe QCD transition from the quark/gluon plasma into hadrons.  Lattice calculations imply that the QCD transition is a ``crossover'' transition at temperature $T_C = 155\MeV$.  Using nucleons, lambdas, and $S$'s as the baryonic degrees of freedom, we show that ``Baryon Statistical Equilibrium'' (BSE)\footnote{Our discussion of Baryon Statistical Equilibrium closely follows the usual discussion of Nuclear Statistical Equilibrium; see e.g., Ref.\ \cite{1990eaun.book.....K}.} is very rapidly established  after the QCD transition, and is maintained only down to a temperature $T\sim 10\MeV$, long before a significant fraction of the baryon number is in dibaryons. In particular, our calculations show that the freeze-out abundance of the $S$'s, which determines the present-day relic abundance, is at most $10^{-11}$ that of nucleons, largely independent of the dibaryon mass and the strength of its interactions.

\section{Baryon Statistical Equilibrium \label{sec:BSE}}

We consider the thermodynamic system of nucleons, lambdas, and dibaryons at temperatures $T<T_C$.  Because other baryons are significantly more massive, only these particles are needed to track the dibaryon abundance.   (Later, we will explicitly show that the only other processes of any importance, those that involve $\Sigma$ and $\Xi$ baryons, can be  ignored at temperatures around freezeout of the dibaryon abundance.) Further, we need not differentiate between neutrons and protons, for which we assume a common mass of $939\MeV$, and $m_\Lambda=1116\MeV$.  For the only unknown in the BSE calculation, we consider the range discussed above, $1860 \MeV < m_S < 1890 \MeV$, where the $S$ is stable or long-lived. 

During the radiation-dominated era the age of the universe and the expansion rate are related by:  $t= 1/2H = 0.301 g_*^{-1/2}\Mpl/T^2 \simeq 0.8\, (T/\mathrm{MeV})^{-2} \s$, where $g_* \simeq 10-20$ at the temperatures of interest counts the effective number of degrees of freedom, which includes photons, electrons, positrons, neutrinos, and smaller numbers of other hadrons. Even just after the QCD transition, the age of the universe is \textit{much} longer than a typical strong-interaction timescale ($10^{-23}\s$), and longer than the timescale for weak decays of hadrons (e.g., the $\Lambda$ lifetime is $2.6\times10^{-10}\s$), and weak interactions more generally (e.g., $e^+ + e^- \longleftrightarrow \nu_i + \bar\nu_i$ and $n + e^+ \longleftrightarrow p + \bar\nu_e$).  The same of course holds for electromagnetic interactions.  In sum, during the period of interest, $T \simeq 155\MeV$ to a few MeV, the constituents of the radiation soup are known and their reactions rapid, which makes equilibrium thermodynamics appropriate.

Recall, there are two types of equilibrium: \textit{kinetic} equilibrium and \textit{chemical} equilibrium.  If a species $i$ is in kinetic equilibrium its phase-space density is $f(E) = g_i\left[\exp\left((E-\mu_i)/T\right)\pm1\right]^{-1}$ where $g_i$ is the number of degrees of freedom, $\mu_i$ is the chemical potential of species $i$, and $+1$ is used for fermions and $-1$ used for bosons. In the nonrelativistic (NR) limit the number density of species $i$ is
\begin{equation}
n_i = g_i \left(\frac{m_iT}{2\pi}\right)^{3/2}e^{-m_i/T}e^{\mu_i/T}\com 
\end{equation}
independent of spin statistics.  As noted above,  strong, electromagnetic, and weak interaction rates are so rapid  that kinetic equilibrium is maintained throughout.

If a reaction $a+b+\cdots \longleftrightarrow \cdots+y+z$ is fast compared to the dynamical timescale for expansion (the age of the universe), chemical equilibrium results, and the sum of chemical potentials in the initial and final states  are equal: $\mu_a+\mu_b+\cdots = \cdots + \mu_y+\mu_z$.  Chemical equilibrium for the process $\Lambda \longleftrightarrow N+\pi+\pi$ enforces $\mu_\Lambda=\mu_N$ since $\mu_\pi=0$ (e.g., $\pi^0 \longleftrightarrow \gamma + \gamma$), while chemical equilibrium for the process $S + \pi \longleftrightarrow \Lambda+\Lambda $ (and other reactions) enforces $\mu_S=2\mu_\Lambda=2\mu_N$.  Using $g_N=4$ (counting neutrons and protons), $g_\Lambda=2$, and $g_S=1$, and assuming all species are NR, the BSE abundances of the $\left\{N,\ \Lambda,\ H\right\}$ system are
\begin{subequations}
\begin{align}
n^\EQ_N(T) & = 4\left(\frac{m_NT}{2\pi}\right)^{3/2}e^{-m_N/T}e^{\mu_N/T} \com \\
n_\Lambda^\EQ(T) & = 2\left(\frac{m_\Lambda T}{2\pi}\right)^{3/2}e^{-m_\Lambda/T}e^{\mu_N/T} = \frac{1}{2}n^\EQ_N(T)\left(\frac{m_\Lambda}{m_N}\right)^{3/2}e^{B_\Lambda/T} \com \\
n_S^\EQ(T) & = \left(\frac{m_ST}{2\pi}\right)^{3/2}e^{-m_S/T}e^{2\mu_N/T} = \frac{1}{16}\left[n^\EQ_N(T)\right]^2 \left(\frac{m_ST}{2\pi}\right)^{3/2}\left(\frac{2\pi}{m_NT}\right)^3e^{B_S/T} \com
\end{align}
\end{subequations}
where  $B_\Lambda \equiv m_N-m_\Lambda = -177\MeV$ and $B_S \equiv 2m_N-m_S=1878 \MeV-m_S$ are the ``baryonic binding energies.''  Note that $B_\Lambda$ is negative (nucleons are a lower baryon energy state), and $B_S$, which is between $18\MeV$ and $-12\MeV$, is positive if $S$ is stable or slightly negative if it is only long-lived.  

Baryon number conservation is expressed as the constancy of baryon number to entropy ratio, $n_B/s =3.9\times 10^{-9}\Omega_Bh^2 \simeq 8.7\times 10^{-11}$, where the entropy density is $s \equiv 2\pi^2 g_* T^3/45$ and $n_B \equiv n_N + n_\Lambda + 2n_S$.

\begin{figure}[t]
\begin{center}
\includegraphics[scale=.7]{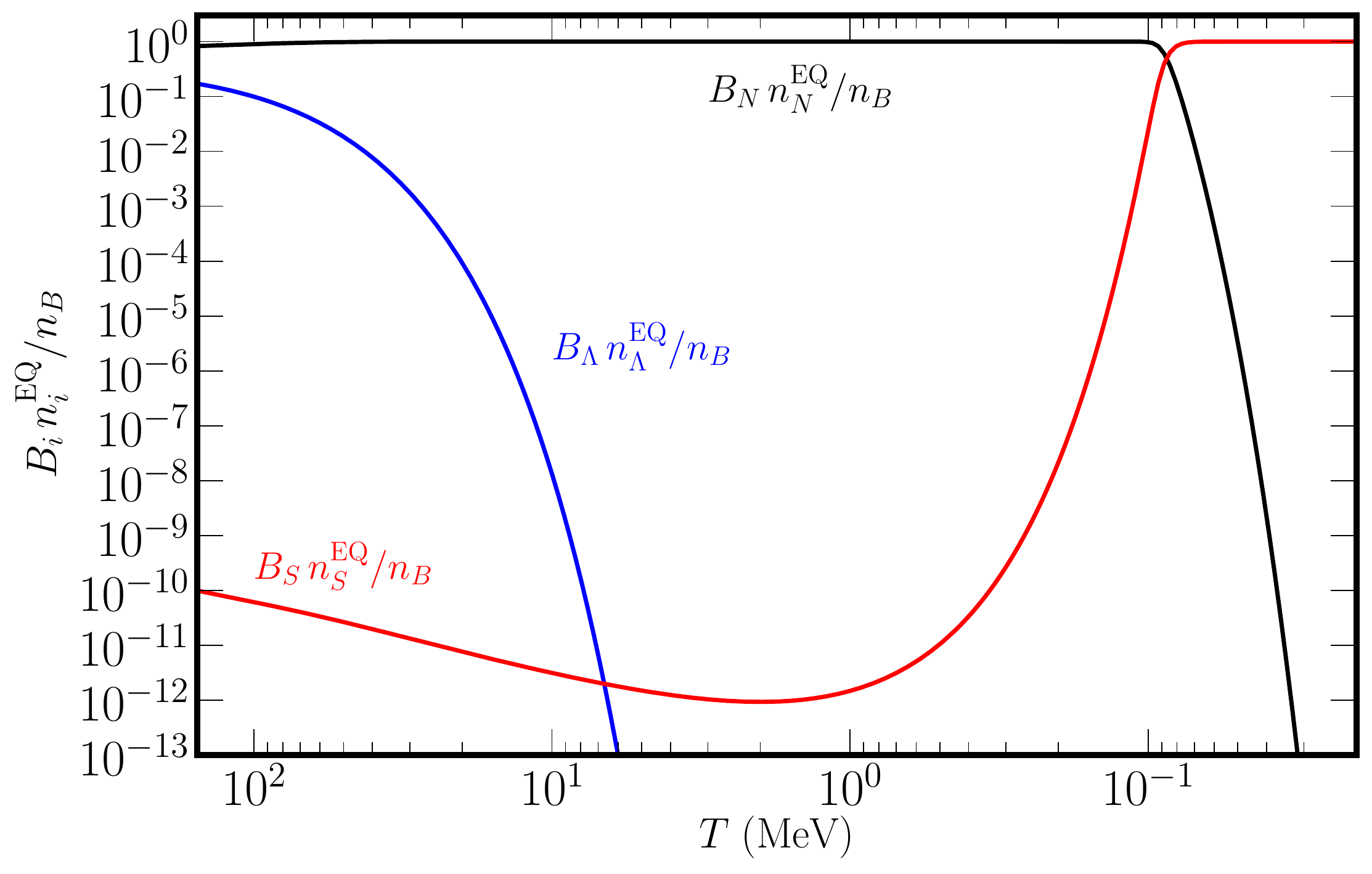}
\caption{\label{fig:BSE} The BSE baryon fractions for the $\left\{N,\Lambda,S\right\}$ system as a function of temperature for $m_S=1875\MeV$.  }
\end{center}
\end{figure}

The BSE baryon fractions for the $\left\{N,\Lambda,S\right\}$ system as a function of temperature assuming $m_S=1875\MeV$ are shown in \fref{fig:BSE}.  There are several important points to note: (1)  The  baryon number carried by $\Lambda$'s is always small and decreases exponentially with temperature because of the $\Lambda$'s negative binding energy.  This is why the heavier baryons can be neglected:  their abundances would be smaller and would decrease faster.  (2)  Assuming $B_S >0$, the baryon fraction carried by $S$'s becomes of order unity at a temperature $T \sim 0.1\MeV$.  While $S$'s are energetically favored for positive $B_S$, owing to the high entropy of the universe ($\eta^{-1} \sim 10^{9}$) $S$'s are not thermodynamically favored until a very low temperature, $T_S \sim B_S/\ln \eta^{-1}$.\footnote{For a similar reason, despite the $7\MeV$ per nucleon binding energy of $^4$He, BBN does not commence until a temperature of order $0.3\MeV$.} (3) In the case of negative $B_S$, $S$'s are never thermodynamically favored and the baryon fraction they carry decreases exponentially in a manner similar to $\Lambda$'s.  (4) The most important result is that for the mass range we consider, $1860 \MeV < m_S < 1890 \MeV$, and for temperatures $10\MeV < T < T_C$, in BSE the baryon fraction carried by $S$'s is between $10^{-11}$ and $10^{-13}$.  This fact makes our predictions for the final abundance of $S$'s insensitive to the mass of the $S$ and its precise cross sections.

Equilibrium thermodynamics is not the entire story and  next we  discuss the freeze out of $S$'s, which determines the dibaryon abundance today.

\section{Freeze out}\label{sec:freezeout}

Equilibrium amongst the three species in our system of baryons, $\left\{N,\Lambda,S\right\}$, only pertains so long as the key reactions that maintain chemical equilibrium are occurring rapidly on the expansion timescale: i.e., reaction rate $\Gamma > H$.  Those key reactions are:  (1) $\Lambda \longleftrightarrow N+\pi$ and (2) $ \Lambda+\Lambda\longleftrightarrow S+ \pi +\pi $.  The first reaction -- a weak process -- regulates the number of $\Lambda$'s through decays and inverse decays.  Since the lifetime of the $\Lambda$ is of the order of $10^{-10}\s$ -- much shorter than the age of the universe at the times of relevance here -- the first reaction ensures that the abundance of $\Lambda$'s closely tracks its equilibrium value.  On the other hand, the second reaction, which regulates the number of $S$'s, cannot keep pace with the expansion at low temperature because of the exponentially decreasing numbers of $\Lambda$'s.  This prevents the dibaryon abundance from increasing and maintaining its BSE value and leads to the small final abundance of $S$'s.

\begin{figure}[t]
\begin{center}
\includegraphics[scale=.7]{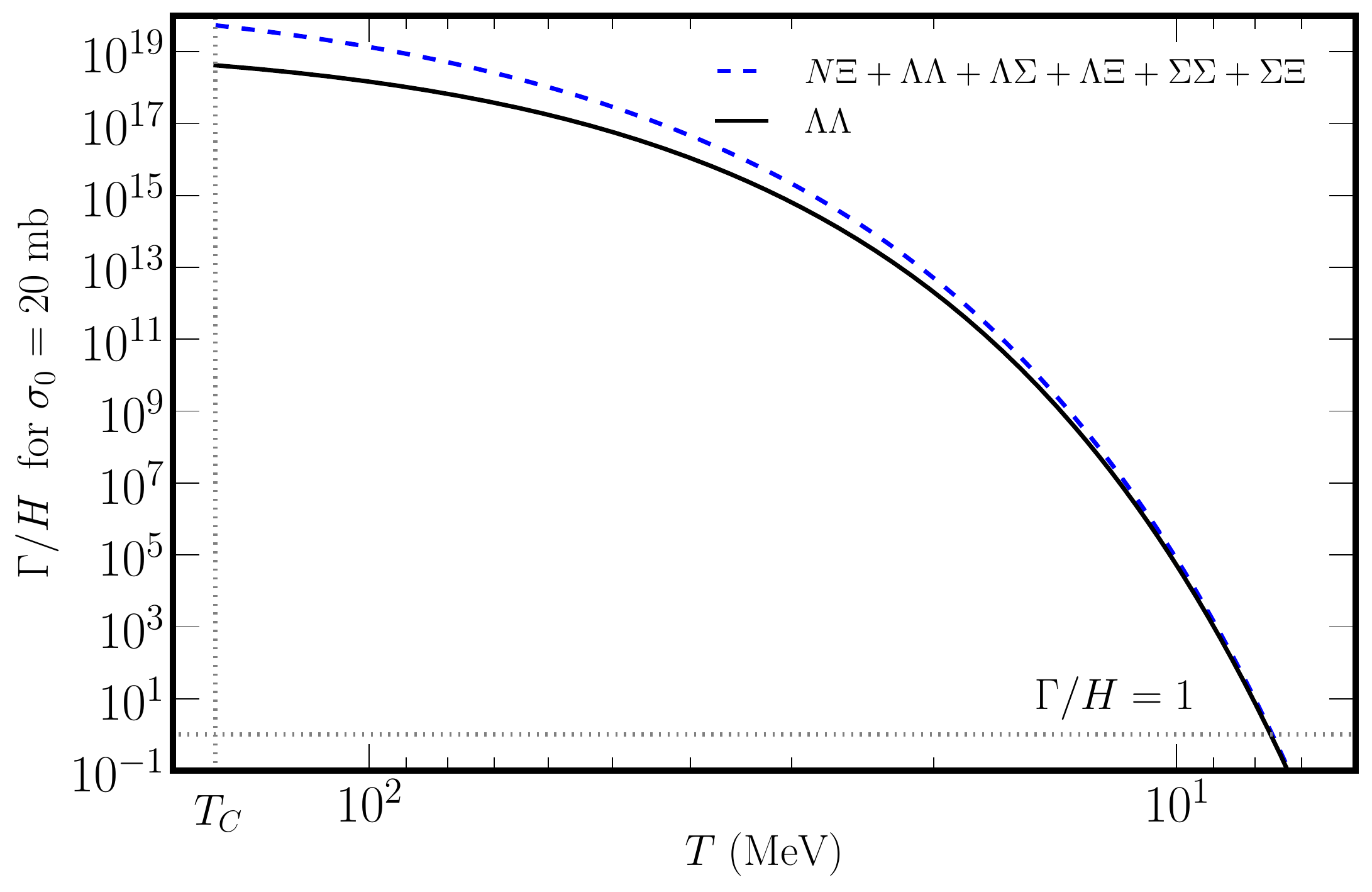}
\centering
\llap{\shortstack{%
	\includegraphics[scale=.35]{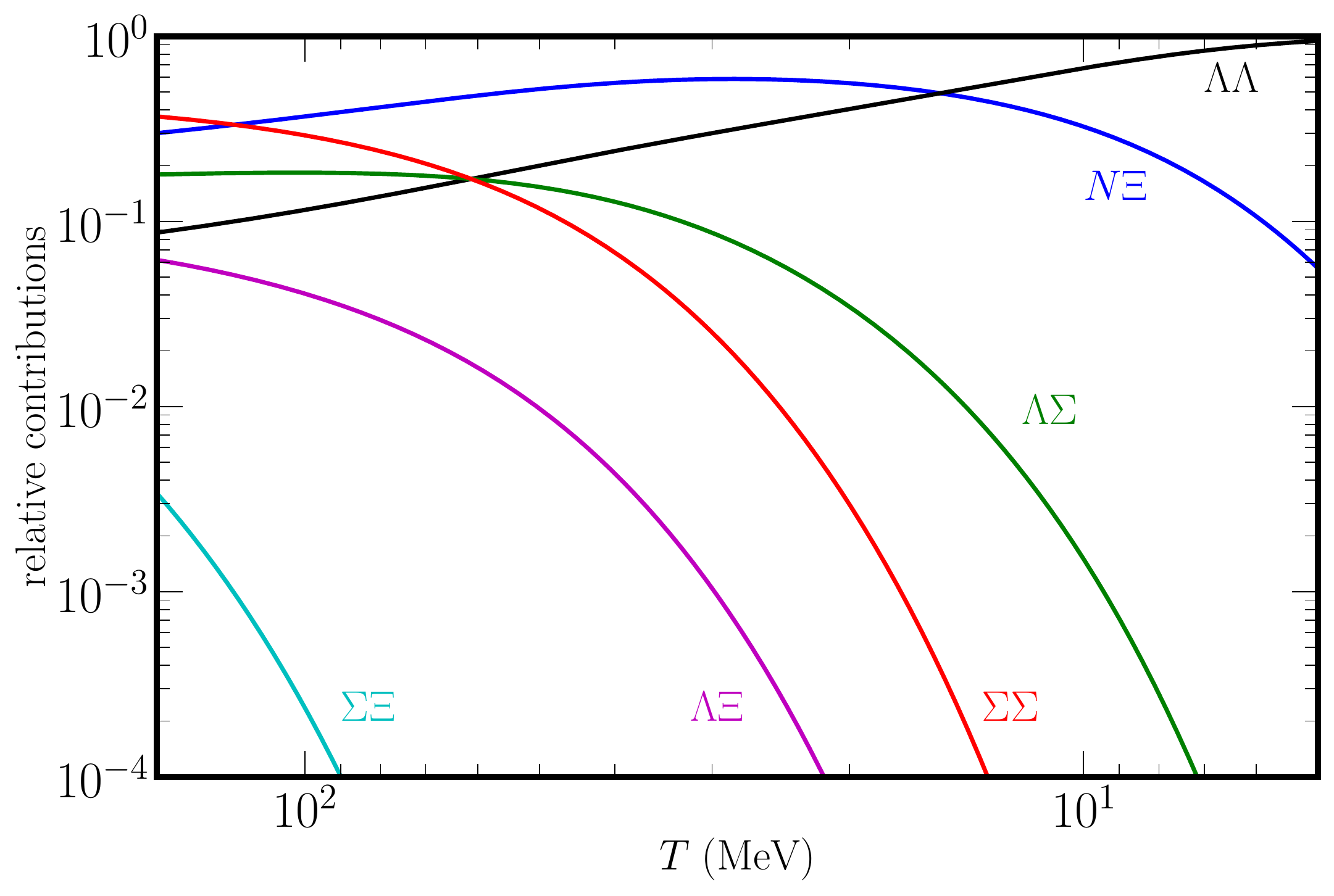}\\
	\rule{0ex}{.67in}%
      }
  \rule{2.1in}{0ex}}
\caption{\label{fig:ratesoverh} The ratio of the reaction rate $\Gamma_{\Lambda \Lambda \rightarrow S\pi\pi}$ to the expansion rate $H$.  Shortly after the QCD transition, the ratio is extremely high, ensuring BSE and a small value for $S$ fraction, and around $8\MeV$ the ratio is unity and the $S$ abundance freezes out.  Shown in the inset are the relative rates for $S$ production through all exothermic processes, i.e., $\Lambda\Lambda$, $\Sigma\Sigma$, $\Sigma\Xi$, $\Lambda\Sigma$, $\Lambda\Xi$ and $\Xi N$ in the initial state.  At high temperature, when $\Gamma /H$ is extremely large all three processes are important, and at freeze out the $\Lambda\Lambda$ process dominates.}
\end{center}
\end{figure}

In principle, three rate equations are needed to follow the evolution of the $\left\{N,\Lambda,S\right\}$ network.  However, because strong and weak interaction rates are so much larger than the expansion rate, the nucleons and $\Lambda$'s will be in chemical equilibrium, and we need only consider the Boltzmann equation for the dibaryons. Using detailed balanced, the equation governing the dibaryon number density can be written as:
\begin{equation}
\dot{n}_S+3Hn_S = - \Gamma_{\Lambda\Lambda\rightarrow S\pi \pi}\left(n_S-n_S^\EQ\right) \per\label{eq:Hdot}
\end{equation}
In \eref{eq:Hdot}, the term $\Gamma_{\Lambda\Lambda\rightarrow S\pi \pi}$ is given by
\begin{equation}
\Gamma_{\Lambda\Lambda\rightarrow S\pi \pi} = \frac{1}{16\pi^2}\frac{1}{T^2}\frac{1}{(m_S/T)^2\,K_2(m_S/T)} 
\int_{4m_\Lambda^2}^\infty ds \left(s-4m_\Lambda^2\right)\sqrt{s}\ \sigma(s)\ K_1(\sqrt{s}/T)\com 
\label{eq:gammall}
\end{equation}
where $K_n(x)$ is the modified Bessel function of the second kind of order $n$, and $\sigma(s)$ is the cross section as a function of $s$ for the process $\Lambda\Lambda \longrightarrow S\pi \pi$.\footnote{The final state $S\pi$ is forbidden in strong interactions by isospin conservation.} This process does not involve a strangeness-changing interaction, so it should proceed via the strong interaction. Absent precise information for $\sigma(s)$, we assume $\sigma(s)=\sigma_0v_\Mol$ in the NR limit, where $v_\Mol$ is the M\o ller velocity. With this choice 
\begin{equation}
\Gamma_{\Lambda\Lambda\rightarrow S\pi \pi} = \sigma_0\,\frac{\left[n^\EQ_\Lambda\right]^2}{n_S^\EQ} = \sigma_0\,4T^3\left(\frac{m_\Lambda}{m_S}\right)^{3/2}\left(\frac{m_\Lambda}{2\pi T}\right)^{3/2} e^{-(2m_\Lambda+m_S)/T} \per
\end{equation}
Further, we  take $\sigma_0 = 1/m_\pi^2=20\ \textrm{mb}$ as the nominal value, and vary $\sigma_0$ from 1 b to 1 pb to illustrate the insensitivity of our results to  precise knowledge for this cross section.  We note that had we included other processes involving $\Sigma$'s and $\Xi$'s that produce $S$'s, e.g., $\Xi N \rightarrow S \pi$, $\Lambda\Sigma\rightarrow S\pi$, or $\Sigma \Sigma \rightarrow S \pi$, the r.h.s of \eref{eq:Hdot} would be replaced by the sum $\sum\limits_{i,j} \Gamma_{i j \rightarrow S X}$ with $\Gamma_{ij}=\sigma_0\, n_i^\mathrm{EQ}n_j^\mathrm{EQ}/n_S^\mathrm{EQ}$.

Shown in \fref{fig:ratesoverh} is the ratio of $\Gamma_{\Lambda \Lambda \rightarrow S \pi\pi}$ to the expansion rate for temperatures between $5\MeV$ and $155\MeV$. Around $T_C = 155\MeV$, when $\Lambda$'s are very abundant, the ratio is greater than $10^{19}$, ensuring that regardless of its initial abundance, the $S$'s will rapidly come into equilibrium.  As the temperature falls, $\Gamma /H$ drops rapidly, because of the exponentially decreasing numbers of $\Lambda$'s, and freezes out (achieves a value of unity) around $T = 8\MeV$.  

Shown in the inset of \fref{fig:ratesoverh}, are the additional $S$ production rates involving $\Sigma$'s and $\Xi$'s (assuming the same cross section normalization).  While these processes can be important around $T_C$, increasing $\Gamma /H$ by about a factor of 3, the final $S$ abundance is determined by the freeze out of $\Lambda \Lambda \rightarrow S \pi \pi$, justifying our simple network of $N$'s, $\Lambda$'s and $S$'s.  We note that $\sigma_0$ would have to be less than about $2\times10^{-46}\cm^2$ (0.2 zeptobarns) for the reactions that regulate $S$'s to be too slow to establish BSE immediately after the QCD transition.

\begin{figure}[t]
\begin{center}
\includegraphics[scale=.7]{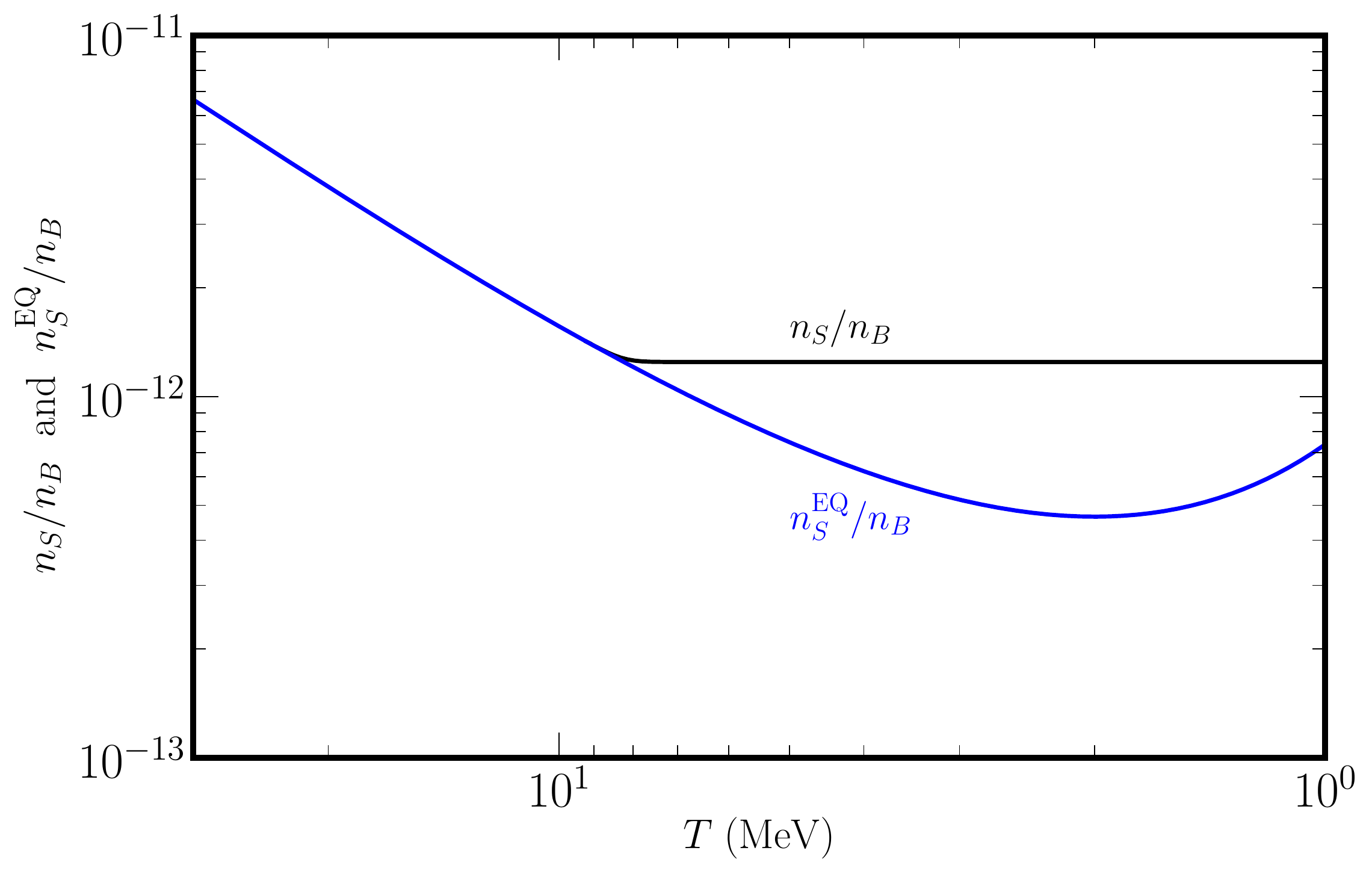}
\caption{\label{fig:freezeout} Results of integrating the $\left\{N,\Lambda,S\right\}$ network for the baryon fraction carried by the $S$ species;  $m_S=1875\MeV$ and $\sigma_0 = 1/m_\pi^2\simeq 20\mathrm{\ mb}$. Note that freeze occurs at a temperature consistent with \fref{fig:ratesoverh}.}
\end{center}
\end{figure}

\begin{figure}[t]
\begin{center}
\includegraphics[scale=.7]{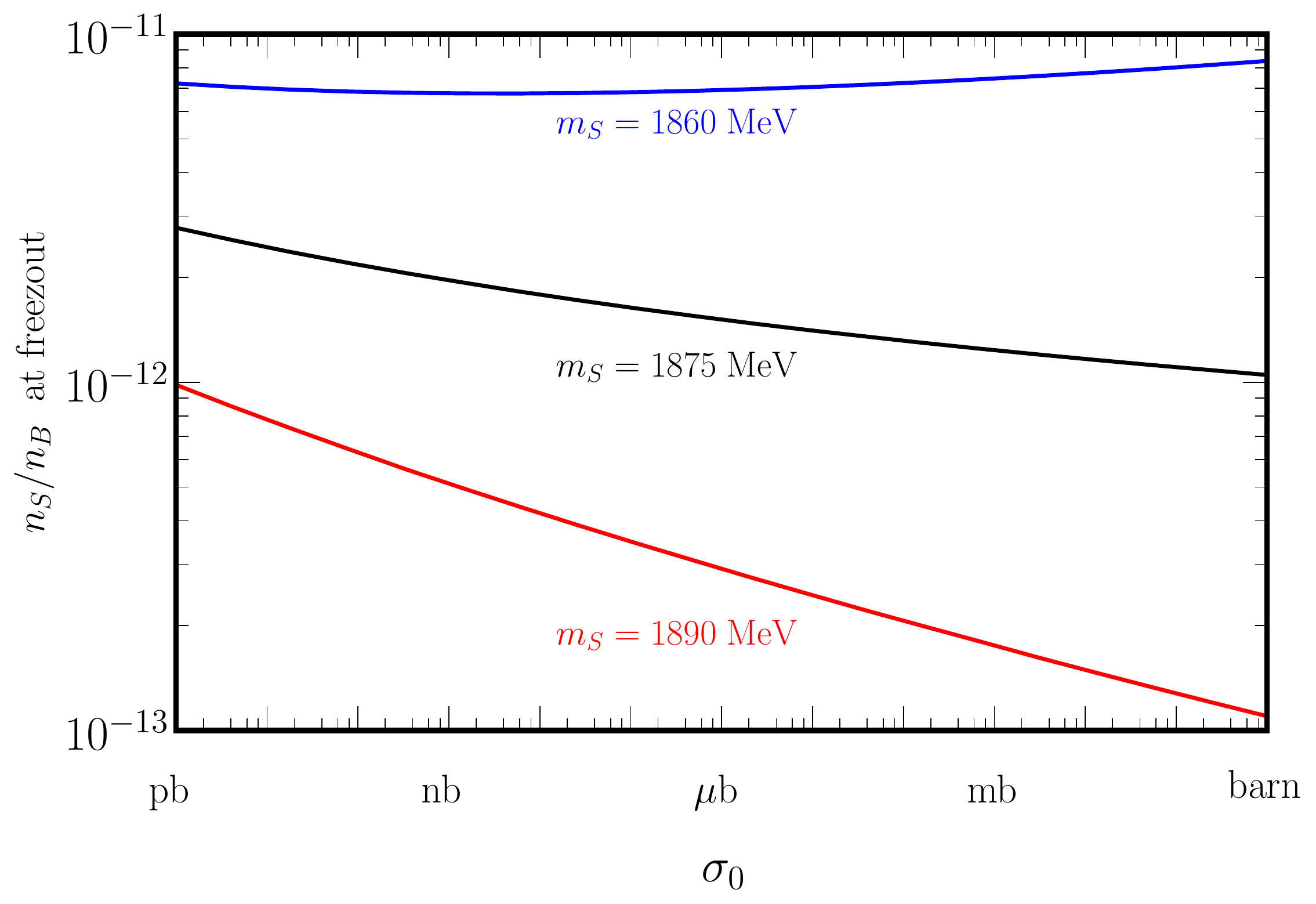}
\caption{\label{fig:results} The final baryon fraction carried by $S$'s as a function of $\sigma_0$ for the mass range of interest.}
\end{center}
\end{figure}

As previously mentioned, the BSE abundance is relatively constant over the range of freeze-out temperatures, at an $S$-abundance to nucleons in the range $10^{-11}$ to $10^{-13}$, and so we expect the final abundance of $S$'s to be in this range.  Fig.~3 shows the results of integrating \eref{eq:Hdot} for an $S$ mass of $1875\MeV$.  As expected, freeze out occurs at a temperature of around $8\MeV$ and an $S$ baryon fraction of about $10^{-12}$.  This freeze out occurs well before BBN, and $S$ production and BBN should not interfere with one another.  Figure \ref{fig:results} shows the final abundance of $S$'s and its insensitivity to $m_S$ and $\sigma_0$.

\section{Conclusions}\label{sec:wrapup}

The production of dibaryons following the QCD transition phase at a temperature of around $155\MeV$ lends itself to an equilibrium thermodynamic treatment because the crucial reaction rates are so rapid compared to the expansion rate of the universe ($\Gamma /H \sim 10^{20}$). In turn, the attainment of equilibrium means that the final abundances are insensitive to the details of how the quark/gluon plasma transitions to hadronic matter.  Using a simple network of nucleons, lambdas, and dibaryons and the key reactions that govern their abundances, we find that the final baryon fraction carried by $S$'s is determined by the BSE value when the reaction $\Lambda \Lambda \rightarrow S \pi\pi $ freezes out.  For the relevant range of $S$ masses ($1860\MeV$ to $1890\MeV$) and a very wide range of cross sections for this process (more than 12 orders-of-magnitude), freeze out occurs between $T \sim 6\MeV$ and $12\MeV$, where the BSE abundance of $S$'s is relatively constant, at a value no larger than $10^{-11}$ that of nucleons.  

As attractive as the idea of dibaryonic dark matter is, we conclude that dibaryons can be but a tiny fraction of the dark matter.  There are two ways to avoid this conclusion; but neither is a easy way out.  

The first way out is if the $S$ is light enough ($m_S \simeq 1200$ MeV) that the BSE abundance is large at freeze-out.  This is the conclusion of Gross et al.\  \cite{Gross:2018ivp}. However, a mass this small is strongly excluded by the arguments given in the \textit{Introduction}.

The second way out is if $S$'s are copiously produced in the QCD transition (e.g., by ``entrainment'' as proposed in \cite{Farrar:2018hac}) and their interactions thereafter are too weak to establish the $S$ in BSE.  In our calculations we assumed that the reactions that regulate the $S$ abundance ($\Lambda\Lambda$, $\Sigma\Sigma$, $\Sigma\Xi$, $\Lambda\Sigma$, $\Lambda\Xi$ and $\Xi N \leftrightarrow S +X$) are strong interactions, with $\sigma_0 = 20$ mb.\footnote{We also assumed $s$-wave reactions.  If the reaction is $p$-wave there is a suppression of $6T/m_\Lambda=0.8$ at $T=T_C$.}  As seen from \fref{fig:ratesoverh}, avoiding BSE just after the QCD transition would require a reduction of about a factor of $10^{20}$ over our assumed cross section.  Farrar argues that while the $S$ has strong interactions with baryons,  a very large suppression in the effective $S$-baryon-baryon coupling can arise due to wavefunction mismatch \cite{Farrar:inprep}.  She needs ten orders-of-magnitude suppression to avoid being ruled out by the experimental constraint arising from doubly-strange hypernuclei, and as mentioned here, 20 orders-of-magnitude  to avoid BSE and our conclusion that dibaryons are not abundant enough to be the dark matter.

Although we conclude that the dibaryon is unlikely to be dark matter, if a stable or long-lived dibaryon does exist they may be an interesting relic, even if the number left over from the early universe is around $10^{-11}$ or so per nucleon.  Given that they are neutral and  have strong interactions, they should bind to nuclei (however, see Ref.\ \cite{Farrar:2003gh} for the argument that the $S$ does \textit{not} bind to nuclei.) The relevant nuclei in the early universe are $p$'s and $^4$He and the formation of $S$-nuclei should occur at a temperature $T \sim BE/\ln \eta^{-1}$ ($BE$ is the binding energy of an $S$ to a nucleus), which is likely to be of order a fraction of an MeV, i.e., around the time of BBN.  (The tiny $S$ abundance ensures that this will not significantly affect BBN.)  Any $S$'s should today masquerade either as an odd, stable form of tritium ($_S^3\textrm{t}$)  or an odd form of $^6$He ($_S^6\textrm{He}$).  By odd, we mean slightly different mass and possibly different nuclear energy levels.  Absent knowledge of their binding energies to $p$'s and $^4$He's, we can't estimate the relative abundances of $_S^3\textrm{t}$ and $_S^6\textrm{He}$.  We do not consider their detectability here.

\begin{acknowledgments}
E.W.K.\ is supported in part by the Department of Energy.  M.S.T.\ is supported at the University of Chicago by the Kavli Institute for Cosmological Physics through grant NSF PHY-1125897 and an endowment from the Kavli Foundation and its founder Fred Kavli.  M.S.T. thanks Gordon Baym and Tetsuo Hatsuda for valuable conversations at the Aspen Center for Physics.  We also acknowledge useful conversations with Alessandro Strumia, Samuel McDermott,  Glennys Farrar and Christopher Hill.
\end{acknowledgments}

\bibliographystyle{apsrev4-1}
\bibliography{S}

\end{document}